\documentclass[aps,prb,amsmath,amssymb,showpacs,superscriptaddress,tighten,longbibliography]{revtex4-2}

\usepackage[english]{babel}

\usepackage[letterpaper,top=2cm,bottom=2cm,left=3cm,right=3cm,marginparwidth=1.75cm]{geometry}

\usepackage{amsmath, bm}
\usepackage{graphicx}
\usepackage[colorlinks=true, allcolors=blue]{hyperref}
\usepackage{siunitx}
\usepackage{booktabs}
\usepackage{xcolor}
\usepackage[normalem]{ulem}

\begin{document}
\title{Modelling the dynamics of the martensitic transformation in $\mathbf{Ni_2MnGa}$ with machine-learning force-fields}
\author{Markus E. Gruner}
\email{Markus.Gruner@uni-due.de}
\affiliation{Faculty of Physics and Center for Nanointegration Duisburg-Essen (CeNIDE), University of Duisburg-Essen, 47048 Duisburg, Germany}
\author{Yulia Klunnikova}
\affiliation{Institute of Materials Science, Functional Materials, Technical University Darmstadt, 64287 Darmstadt, Germany}
\author{Mike J. Bruckhoff}
\affiliation{Faculty of Physics and Center for Nanointegration Duisburg-Essen (CeNIDE), University of Duisburg-Essen, 47048 Duisburg, Germany}
\author{Konstatin P. Skokov}
\affiliation{Institute of Materials Science, Functional Materials, Technical University Darmstadt, 64287 Darmstadt, Germany}
\author{Olga N. Miroshkina}
\affiliation{Faculty of Physics and Center for Nanointegration Duisburg-Essen (CeNIDE), University of Duisburg-Essen, 47048 Duisburg, Germany}

\newcommand{\NiMnGa}{$\mathrm{Ni_2MnGa}$}
\newcommand{\MLFF}{machine-learning force field}
\newcommand{\DFT}{density functional theory}
\newcommand{\MD}{molecular dynamics}
\newcommand{\TB}{twin boundary}
\newcommand{\NM}{non-modulated}
\newcommand{\LTO}{L2$_1$} %
\newcommand{\LONE}{L1$_0$} %
\newcommand{\meVfU}{meV/f.u.}

\begin{abstract}
The martensitic transformations in stoichiometric $\mathrm{Ni_2MnGa}$ and related Heusler alloys are governed by a delicate interplay between electronic structure, magnetism, and lattice dynamics. While density functional theory (DFT) has successfully mapped the energy landscape of modulated martensites, it remains computationally prohibitive for capturing finite-temperature dynamics across relevant length and time scales. Here, we bridge this gap by constructing DFT-trained machine-learning force-fields (ML-FF) interpolated from ab initio molecular dynamics. 
We demonstrate that ML-FF trained on DFT data can reproduce key features of $\mathrm{Ni_2MnGa}$, including the soft TA$_2$ phonon and a preference for nanotwinned martensites. Using classical molecular dynamics simulations, we explore the impact of adaptive nanotwinning on the austenite$\leftrightarrow$martensite transition, which facilitates the dynamics of both, forward and backward transformation. We demonstrate, that the negative twin boundary energy along the entire transformations path predicted by DFT for $T=0$, supports the formation of ordered modulated martensites at finite temperatures. Our results pave the way for large-scale simulations of magnetic shape memory alloys with near-DFT accuracy.
\end{abstract}

\maketitle

\section{Introduction}
The entire family of Ni-Mn-based Heusler compounds shares similar phase diagrams with unique features that make them suitable for functional applications, in particular the magnetic shape memory effect or magnetocaloric applications \cite{cn:Sozinov02,cn:Kainuma06,cn:Liu12,cn:Umetsu16}. Linked to the functional behavior is a martensitic phase transition, which is characterized by the appearance of similar modulated martensitic structures in all these compounds. Common to most of these systems are the non-modulated (NM) tetragonal L$1_0$ structure, the five-fold and seven-fold modulated monoclinic phases 10M and 14M, and another orthorhombic structure called 4O~\cite{cn:Acet11Handbook,cn:Xu13,cn:Xu16}, as established by high-resolution electron microscopy and diffraction studies~\cite{Chernenko1998,cn:Pons00,Pons2005,Lanska2004}. 
Among this family, $\mathrm{Ni_2MnGa}$ stands out because the martensitic phase transition can be observed in the stoichiometric compound at an electron-to-atom ratio of $e/a = 7.5$. This simplifies modelling considerably and makes $\mathrm{Ni_2MnGa}$ an ideal model system to obtain a thorough understanding of the underlying physics before addressing the more complex off-stoichiometric and multi-component systems. 

In the stoichiometric compound, the parent phase is the cubic L$2_1$\ structure with ferromagnetic order of the Mn atoms and induced moments on Ni. With decreasing temperature, it undergoes a transformation into a (pseudo-cubic) 3-fold modulated intermediate premartensite. Depending on composition, the final low temperature martensite can be a -- often incommensurate --  5-fold or 7-fold modulated (pseudo-orthorhombic or monoclinic) structure or eventually also a non-modulated tetragonal L$1_0$ phase~\cite{Kokorin1996,Kokorin2001}. In addition, intermartensitic transformations between the modulated phases have been observed at low temperatures~\cite{cn:Segui05,Chernenko2005b}. 
Recently, it was reported that the modulation period can also change with temperature or external stress~\cite{cn:Vertat21,cn:Vinogradova24}.
The transition %
has been associated with a soft transverse acoustic (TA$_2$) phonon mode along the $[110]$ direction~\cite{Manosa2001,cn:Ener12}.
The condensation of this soft mode has been interpreted as the formation of a premartensitic intermediate phase~\cite{Kokorin1996,Kokorin2001}. This has been traced back early to a Kohn anomaly arising from nesting of sheets of the Fermi surface, observed in theory and experiment~\cite{cn:Velikokhatnyi99,cn:Lee02,cn:Bungaro03,cn:Haynes12,cn:Bhattacharya23}. 
This gives rise to an electron-phonon contribution to the dynamical matrix, which leads to a gain in energy for lattice waves with a matching reciprocal vector. 
The relationship between incommensurate sinusoidal wave-like modulation~\cite{cn:Righi07,cn:Righi08}
as expected from a condensed phonon mode and a discrete nanotwinning picture, which arises from the concept of adaptive martensite~\cite{cn:Khachaturyan91}, 
has long been the subject of debate. Within this framework, the modulated 10M and 14M structures are not regarded as independent thermodynamic phases but rather as nanoscale microstructures of the non-modulated L$1_0$ martensite, 
in which atomically sharp twin boundaries are introduced at the unit-cell level, minimizing the elastic energy associated with the austenite--martensite habit plane~\cite{cn:Kaufmann10,cn:Niemann12}. 
Recent high-resolution diffraction studies reveal that the 10M modulation is strongly anharmonic, with significant high-harmonic components, causing the displacement profile to approach a zig-zag character~\cite{cn:Vertat24,cn:Vertat26}, 
which might be understood as the continuous modulation wave relaxing locally towards non-modulated tetragonal distortions, bridging the wave and nanotwinning descriptions.

Density functional theory (DFT), which offers parameter-free access to ground-state properties, 
such as electronic and magnetic structure, lattice parameters, atomic positions, forces, and phonon spectra, simultaneously and independently of experimental input, 
has been applied to reveal the interplay between electronic and lattice structure with tremendous success~\cite{cn:Ayuela02,cn:Enkovaara03,cn:Entel13,cn:Gruner14}.
Within the DFT framework, the parent L$2_1$\ phase of \NiMnGa\ is predicted to have a ferromagnetic ground state with large local moments on Mn and smaller induced moments on Ni~\cite{cn:Ayuela02,cn:Enkovaara03} and RKKY-type long-ranged oscillating exchange coupling~\cite{cn:Sasioglu04,cn:Buchelnikov10}. 
The total energy of \NiMnGa\ as a function of the tetragonal distortion along the Bain path reveals a shallow minimum for the cubic austenite with L$2_1$\ ordering and, across a tiny energy barrier, a deeper minimum at a tetragonal distortion of $c/a \approx 1.25$, which we identify as the non-modulated L$1_0$\ martensite (NM). However, the actual transformation path in the material is not a simple Bain path but involves additional modulations.
Since DFT calculations are performed at $T=0$, the soft TA$_2$ mode becomes imaginary; this means that the corresponding sinusoidal modulations lead to an energy gain with respect to the parent phase. 
Full optimization of lattice parameters, angles, and atomic positions further decreases the energy of these modulated structures. 
In the end, strongly anharmonic 10M, 14M, and 4O modulations are found to compete closely in energy with the non-modulated L$1_0$ martensite~\cite{cn:Zeleny16,cn:Gruner18,cn:Zeleny18}. The energy differences are of the order of only a few \meVfU, which is consistent with the close competition observed in the experimental phase diagrams. 
Moderate changes to the exchange-correlation functional in terms of a small Hubbard U correction for the Mn sites can stabilize the 10M ground state and improve the agreement with experimental data in terms of lattice parameters, elastic constants, and magnetic anisotropy~\cite{cn:Zeleny21}. 
More recently, efforts were dedicated to the accurate determination of elastic constants and the generalized planar fault energies (GPFE)~\cite{cn:Sestak26}. 

While DFT has proven efficient at identifying peculiar ground-state features that are potentially connected to the functional properties of Ni$_2$MnGa, 
their relevance at finite temperatures remains unproven. 
The phase sequence has been reproduced accurately using thermodynamic calculations in the quasiharmonic approximation~\cite{cn:Uijttewaal09,cn:Dutta16} quite early, 
but many questions remain open regarding the extension of the DFT results to finite temperatures, large systems, and longer time scales. 
This includes the origin of the peculiar shift of the modulation vector, GPFE at finite temperatures, or the role of the phonon instability for the functional properties of this material, which can only be revealed by dynamic simulations at $T>0$. 
First-principles approaches are extremely time-consuming for such cases. Consequently, mainly quasi-static DFT investigations have been available to date~\cite{cn:Gruner08,cn:Zeleny18,cn:Sestak26}. 

In turn, the relevant length and time scales become accessible in the framework of classical molecular dynamics (MD), which often relies on semi-empirical analytic interatomic potentials developed for specific applications and materials. Such potentials are available for several conventional shape-memory systems (e.g.\ NiAl, FeNi, NiTi)~\cite{cn:Becquart93,cn:Meyer98,cn:Guo07,cn:Ko15,cn:Chowdhury17} at the price of some loss of accuracy.
However, for the complex physics of the magnetic Ni-Mn-based Heusler compounds, no empirical potentials of sufficient accuracy exist; their development is generally difficult and tedious, with uncertain success. 

In recent years, a new and powerful approach has become available: classical MD with machine-learning force fields (ML-FF) trained on DFT data. The promise is to achieve the speed of classical MD with near-DFT accuracy (see, e.g.,~\cite{cn:Deringer19}). 
The construction of such a force field often involves three main steps: (i)~generating a reference database from ab initio calculations for several thousand structures, 
selected from ab initio MD runs. (ii)~A suitable representation of the local atomic environment, for instance using the Smooth Overlap of Atomic Positions (SOAP), which encodes 2-body, 3-body, and many-body information from all neighbors within a cutoff radius. (iii)~Fitting using a regression method to interpolate forces and energies on the potential energy surface.
On-the-fly ML-FF training based on Bayesian error estimates, as implemented in VASP, allows the database to be expanded adaptively as the simulation proceeds~\cite{cn:Jinnouchi19PRL,cn:Jinnouchi19PRB}. 

In the following, we construct two versions of an ML-FF for \NiMnGa\ that differ mainly in the inclusion of nano-twinned intermediate states in the training set. This allows for a controlled comparison that can isolate the impact of adaptive nanotwinning on the energy landscape, phonon stability, and transformation kinetics. 
We benchmark both versions against DFT and use them to perform large-scale classical MD simulations of the forward and backward martensitic transformations, demonstrating that the proper description of intermediate martensites is essential for reproducing the transformation pathways.
\section{Computational Details}
\label{sec:methods}

\subsection{DFT calculations}
\label{subsec:dft_details}
Our first-principles calculations were carried out in the framework of density functional theory using the Vienna Ab initio
Simulation Package (VASP)~\cite{VASP1,VASP2} version 6.5.1 and 6.6.0. The code employs a plane-wave basis in combination with the projector augmented-wave approach to describe the interaction between valence and core electrons.~\cite{cn:PAW} 
We used PAW potentials with a valence of $3d^{10}$ for Ni, $3d^6$ for Mn, and $4s^24p^1$ for Ga, and the generalized gradient approximation (GGA) in the form of
Perdew-Burke-Ernzerhof (PBE), with a cutoff $E_{\rm cut}=352\,$eV together with
Methfessel-Paxton smearing~\cite{cn:Methfessel89} with a width of 0.1~eV. 
 
To keep the complexity of the training set tractable, we restricted the calculations to stoichiometric compositions with complete ferromagnetic order. For these conditions, previous DFT investigations predict a 14M modulated ground state and a martensitic transition~\cite{cn:Dutta16,cn:Uijttewaal09}, albeit at temperatures different from experiment unless magnetic disorder is taken into account. 
The DFT calculations are carried out on the Born--Oppenheimer surface. 

Structural optimizations included full relaxation of lattice parameters,
angles, and internal atomic positions unless stated otherwise. 
Phonon frequencies were obtained from the dynamical matrix evaluated with the finite-difference method using the phonopy code~\cite{cn:Phonopy1,cn:Phonopy2}. 
Temperature-dependent free energies were computed in the harmonic approximation, as, e.g., described in \cite{cn:Sokolovskiy19}.

\subsection{Machine-learning force field}
\label{subsec:mlff_details}
During training, ab initio MD was carried out on supercells containing 128--160 atoms, with 48--64 k-points in the full Brillouin zone. 
The ML-FF was constructed following the three-step procedure outlined above, using on-the-fly Bayesian-error-based training as provided in the VASP code (version 6.6.0)~\cite{cn:Jinnouchi19PRB}, 
and essentially follows the procedures described in Refs.~\cite{cn:Liu21,cn:Verdi21}. Radial and angular descriptors were employed with a cutoff radius of 8~\AA{} and 5~\AA{}, respectively, with Gaussian broadening of 0.5~\AA{}, and 12 and 8 basis functions for the radial and angular distributions, respectively, with angular expansion up to $l_{\max} = 3$. 
The mapping from descriptors to energies and forces was learned by
regularized kernel ridge regression. The local atomic environments were represented within the linear regression model as a polynomial kernel of degree 4.

ML-FF v1 serves as the baseline model, trained exclusively on non-modulated structures without explicit treatment of nanotwinned intermediate states along the Bain path. Here, supercells of ferromagnetic L$2_1$ austenite, 6M and 10M pre-martensite, 4O, and non-modulated L$1_0$ martensite with different $c/a$ ratios (1.05, 1.15, and 1.25) were used as training sets and heated from 5~K to typically 750~K with a fast heating rate of $2\ldots 5 \times 10^{14}$~K/s. 
For the cubic $L2_1$ austenite structure, we added further {\em ab initio} MD data up to $2000\,K$. 
We propagated the simulation cell with {\em ab initio} MD in an $NpT$ ensemble using DFT and ML-FF with a time step of 1--2~fs. 
The $NpT$ ensemble was realized with a Langevin thermostat in combination with the Parinello--Rahman scheme, 
with Langevin friction coefficients of $25\,{\rm ps}^{-1}$ for Ni, $12\,{\rm ps}^{-1}$ for Mn, and $7\,{\rm ps}^{-1}$ for Ga. 
The friction coefficient for the lattice degree of freedom was $15\,{\rm ps}^{-1}$, 
and the fictitious lattice mass was reduced to $25\,$amu to establish a fast response to fluctuations. 
DFT calculations were only carried out on-the-fly after the Bayesian error surpassed a critical level, which substantially speeds up the sampling of relevant configurations in the learning phase and resulted in a total of 1703 configurations for the training process.
The radial and angular descriptors were employed with a cutoff radius of 8~\AA{} and 5~\AA{}, respectively, together with Gaussian broadening of 0.5~\AA{} for both. 12 and 8 basis functions were used for the radial and angular descriptors, respectively, with angular expansion up to $l_{\max} = 3$. 

ML-FF v2 extends the training set to 3074 reference configurations by explicitly adding nanotwinned intermediate 4O states along the Bain path. We also added further premartensite and 4O configurations close to $T=0$, obtained from geometric optimization procedures. This extension is designed to capture the soft-mode instability and the energetic preference for nanotwinned configurations along the entire Bain path identified in DFT~\cite{cn:Gruner18}. The difference between v1 and v2 thus provides a controlled assessment in how far the transformation dynamics depends on the description of intermediate nanotwins.
In the latter case, we heated nanotwinned 4O configurations 
generated from non-modulated building blocks with an initial $c/a$ of 1.015, 1.06, 1.11, and 1.21, then heated those configurations to $T=750\,$K, 
keeping the cell shape fixed (but allowing the volume to adapt). 
For ML-FF v2, the radial and angular descriptors were employed with a cutoff radius of 8~\AA{} and 5~\AA{}, respectively, Gaussian broadening of 0.3~\AA{}, and 12 basis functions for both radial and angular descriptors, with an angular expansion up to $l_{\max} = 3$. 

\subsection{Classical molecular dynamics}
\label{subsec:md_details}

After refitting, classical MD simulations were performed with a time step of 2\,fs in the $NpT$ ensemble with the ML-FF at ambient pressure, again using the
Langevin thermostat and the Parinello--Rahman barostat, as
implemented in VASP. For a more realistic response, the fictitious lattice mass was restored to the default value of $1000\,$amu. 
In this way, we simulated systems between approximately 4000 and
10\,000 atoms.

To simulate the ultrafast heating process, the lattice was first equilibrated for 10\,ps at $T=150$~K, then heated with a linear rate from  $T=150$~K to $T=550$~K within
$\Delta t = 2$~ps, and subsequently held at 550~K for an additional
10~ps.
Subsequent cooling and slow heating protocols were carried out with the same parameters but slower heating and cooling rates, as indicated in the text.

\section{Results}

\subsection{Benchmarking of the force-fields} 
\subsubsection{Energy landscape at $T=0$}
\begin{figure}[htbp]
\centering
 \includegraphics[width=0.45\columnwidth]{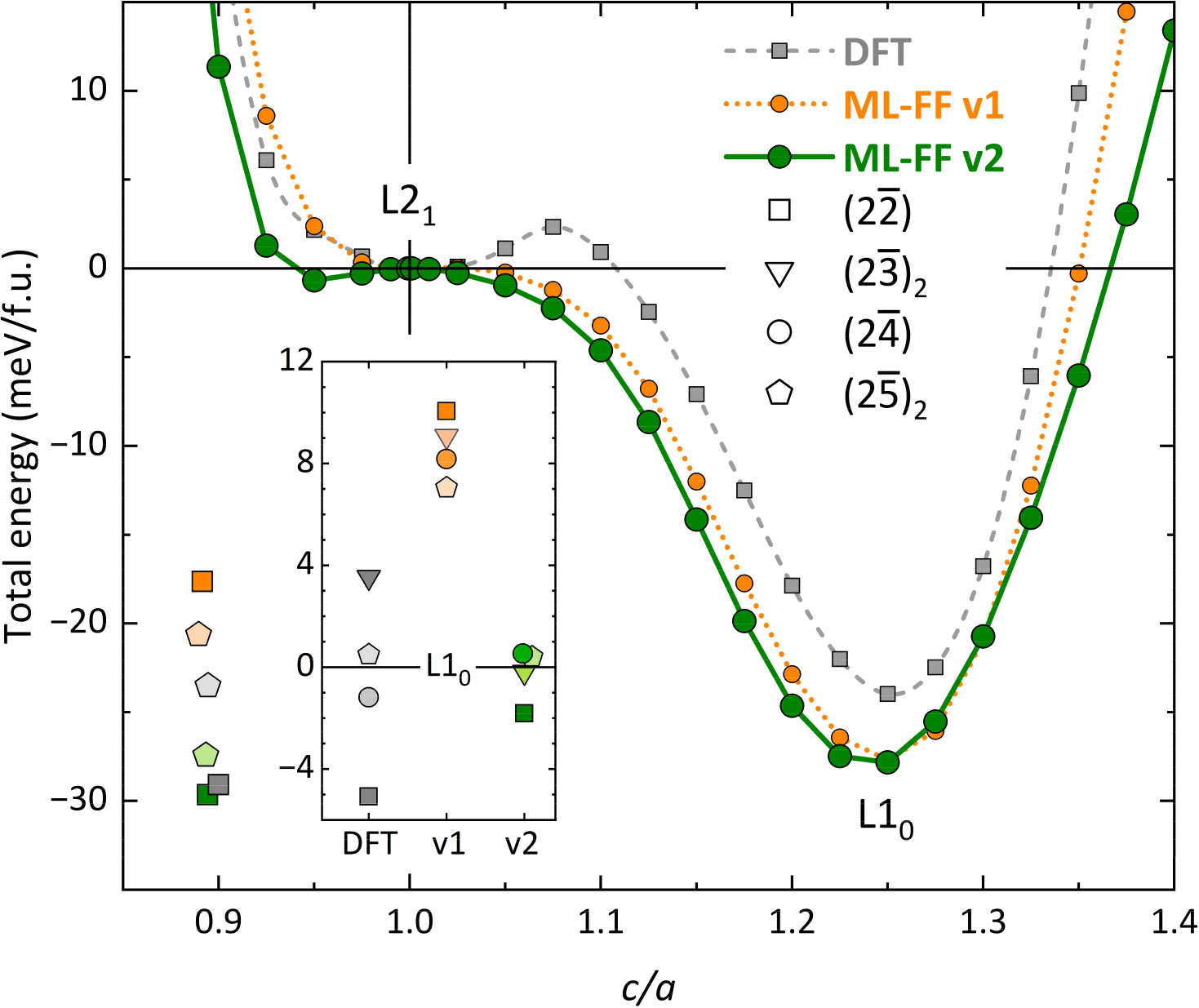}
\caption{Total energy per formula unit as a function of $c/a$ along the Bain path as a benchmark for the ML-FF against DFT. Green color refers to the predictions of ML-FF v2, 
orange to ML-FF v1 and grey to the DFT reference calculated with a technical setup 
comparable to the training set. Energies are plotted relative to the L2$_1$ austenite. Isolated symbols indicate the relative energies of selected modulated structures, denoted in Zhadanov notation: 14M=$(2\overline{5})_2$, 6M=$(2\overline{4})$, 10M=$(2\overline{3})_2$ and 4O=$(2\overline{2})$.  
The inset shows energies relative to the L1$_0$ non-modulated martensite. }
\label{fig:eca}
\end{figure}
As a first benchmark, we compare the total energy per formula unit obtained with the ML-FF against DFT reference data along the Bain path. The latter was generated with a similar technical setup (in terms of valence states, cutoff energy, and k-point density) as the training data. 
For both ML-FF versions, the agreement is within a few \meVfU ~along the Bain path (see Fig.~\ref{fig:eca}). 
In particular, the L$1_0$ minimum is slightly lower than that of the austenite, and the small barrier around $c/a=1.07$ is not reproduced. 
Furthermore, for ML-FF v2 the L$2_1$ austenite is represented by a saddle point with a shallow minimum around $c/a=0.95$ (this part of the Bain path was not included in the training set). 
However, a similar shape for $E(c/a)$ has been observed in DFT calculations of other Ni-Mn-based Heusler compounds~\cite{cn:Siewert11,cn:Entel13}. 
We also compared the nanotwinned structures
with a thickness of two layers for one twin and variable width ($n=2,3,4,5$) for the other, corresponding to 4O, 10M, 6M, and 14M, respectively. For both ML-FF versions, their energies are close and substantially lower than that of the L$2_1$ austenite. 
For ML-FF v2, we obtain a stable 4O ground state as predicted from DFT (cf.\ inset in Fig.~\ref{fig:eca}). This agreement for the modulated structures is remarkable, as only the 4O has been part of the training set. In particular, it is predicted with an accuracy comparable to or better than that of the 4O structure, 
indicating that the ML-FF captures the essential physics of the modulated martensites, supporting the interpretation of these phases 
as nanotwinned variants of the L$1_0$ martensite -- at least for the DFT results.

\begin{figure}[htbp]
\centering
\includegraphics[width=0.35\columnwidth]{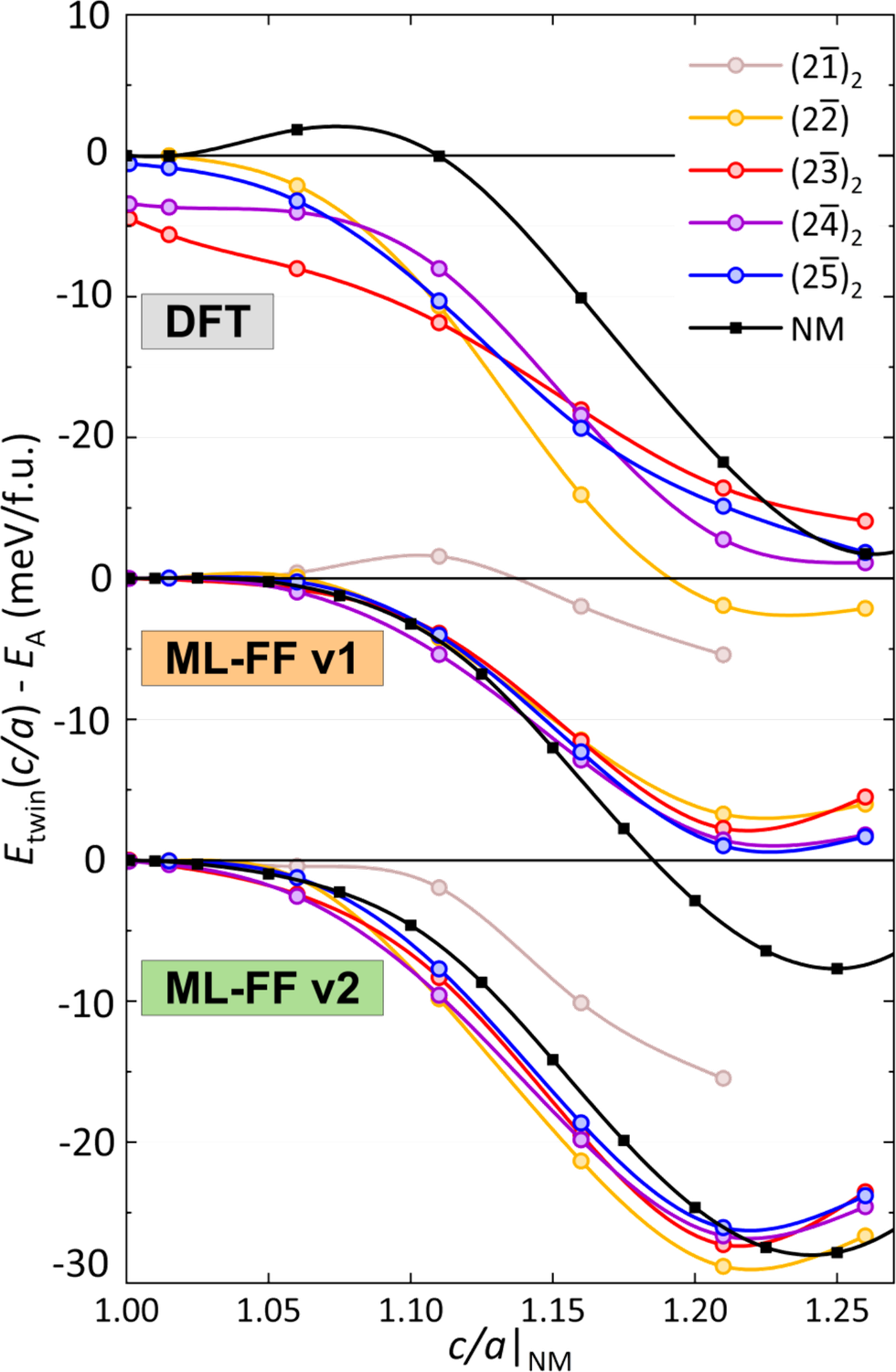}
\caption{ 
Comparison of the total energy of intermediate nanotwinned martensites constructed from non-modulated building 
blocks with fixed $c/a|_{\rm NM}$ along the Bain path (colored lines) compared to the non-modulated L1$_0$ martensite (black line). All structures were relaxed with fixed cell shape. Top panel: DFT reference (taken from \cite{cn:Gruner18}), center and bottom panel: Predictions from ML-FF v1 and v2.
}
\label{fig:etwin}
\end{figure}

To shed more light on the evolution of nanotwinned martensites from cubic austenite, we construct intermediate structures from a stacking of non-modulated L$1_0$ building blocks with a given $c/a$ along the Bain path. As suggested by the concept of adaptive martensite,~\cite{cn:Khachaturyan91,cn:Kauffmann11} we stacked two $[1\,0\,1]$ layers 
with the $c$-axis oriented in one direction and $n$ layers in the other, separated by an atomically sharp interface.~\cite{cn:Gruner18} 
The structures were allowed to relax with the cell shape fixed.
In DFT, this resulted in a downhill energy path towards the modulated martensites, 
with energies systematically lower than their constituent building blocks. In the above picture, this is equivalent to negative twin boundary energies $\gamma_{\rm TB}$ for all these structures. In a previous work, this was traced back to the efficient removal of frustrated antiferromagnetic exchange interactions at the interfaces in two-layered twins,~\cite{cn:Gruner18} explaining the particular stability of the 4O widely observed in DFT~\cite{cn:Zeleny16,cn:Zeleny18}.
The comparison between v1 and v2 in Fig.\ \ref{fig:etwin} demonstrates the impact of the training set on the twin boundary energetics. ML-FF v1 reproduces the vanishing nanotwin boundary energy for small and intermediate tetragonal distortions (i.e., $\gamma_{\rm TB}\approx 0$). Here, the energy curves of the twinned martensites lie on top of the non-modulated martensite curve, indicating no energy cost or benefit for creating a twin interface. However, for larger distortions the twinned structures lie systematically higher than the untwinned L$1_0$, in particular close to their equilibrium distortion. This discrepancy with DFT that is resolved by ML-FF v2, where the explicit inclusion of nanotwinned intermediate states in the training set recovers the negative $\gamma_{\rm TB}$ over a wider range of $c/a|_{\rm NM}$. 
Nevertheless, $\gamma_{\rm TB}\approx 0$ for a wide range of tetragonal distortions has been identified as a salient feature 
of other magnetic shape memory alloys such as Fe$_{70}$Pd$_{30}$~\cite{cn:Weiss11PRL}. Thus, comparing with ML-FF v1 may be helpful to understand the impact of negative twin boundary energies.

We further benchmarked ML-FF v2 by computing the free energy from the quasi-harmonic approximation. The results show excellent agreement with previous DFT calculations,~\cite{cn:Uijttewaal09} confirming that the force field correctly reproduces the thermodynamic stability of the various phases.

\subsubsection{Phonon dispersion and free energy}
\begin{figure}[htbp]
\centering
\includegraphics[width=0.9\columnwidth]{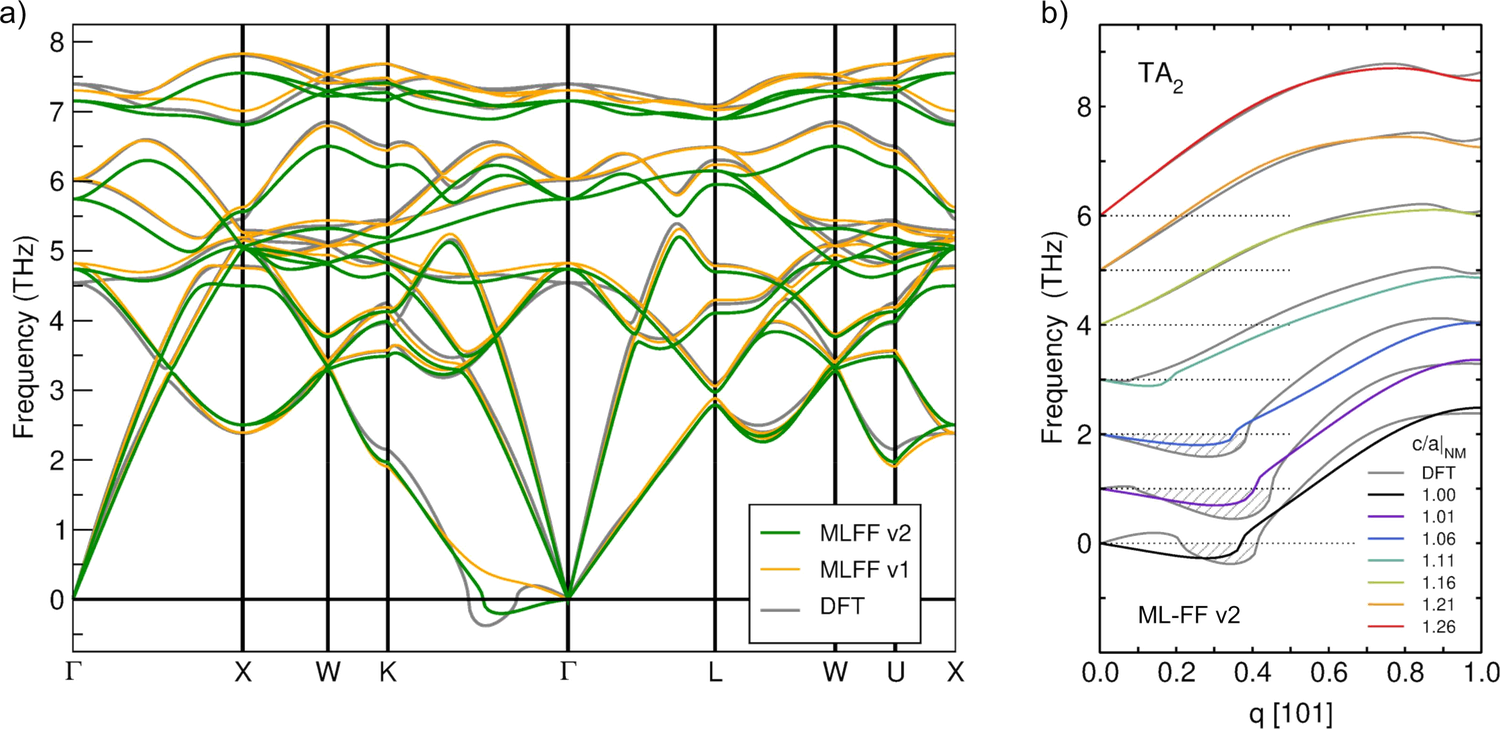}
\caption{
a) Comparison of phonon band structures of cubic L2$_1$ austenite calculated with both versions of ML-FF (orange lines for v1, green lines for v2) and DFT (grey lines). 
Apart from a slight shift in frequency, all modes have nearly the same dispersion. Main difference is the soft TA$_2$ mode along $\Gamma$-K in [101] direction. The instability with imaginary frequencies obtained from DFT is only reproduced by ML-FF v2. b) Soft TA$_2$ mode along [101] calculated for non-modulated tetragonal structures with different $c/a|_{\rm NM}$. 
For visibility, individual curves are shifted vertically by a constant offset. Grey lines denote the DFT reference taken from Ref.~\cite{cn:Gruner18}. 
}
\label{fig:phonons}
\end{figure}
As a benchmark for the validity of the forces, 
the phonon band structure of L$2_1$ austenite was calculated with DFT and both ML-FF versions at the same volume (see Fig.~\ref{fig:phonons}a). 
Both ML-FF versions show excellent overall agreement with the DFT reference for the entire dispersion, with ML-FF v2 appearing slightly softer, which we ascribe to the slight differences in equilibrium volumes between ML-FF and DFT.
However, the soft phonon mode along the $\Gamma$--K direction has effectively disappeared in ML-FF v1. It only survives as a slight dip in the dispersion, 
 meaning that a stable premartensitic modulation is not expected at $T=0$ with the baseline force field. 
In ML-FF v2, where nanotwinned intermediate states are explicitly included in the training set, the imaginary phonon along $\Gamma$--K is restored. As in the DFT case, the instability survives in the non-modulated intermediate tetragonal martensites up to $c/a=1.1$. Fig.~\ref{fig:phonons}b) shows a remarkable agreement for the TA$_2$ branch in the $[1\,0\,1]$ direction between austenite at $c/a=1.0$ and stable martensite at $c/a=1.26$.
 The comparison between v1 and v2 thus confirms that the particular of inclusion intermediate nanotwinned martensites are the decisive ingredient for reproducing the soft-mode instability -- rather than a general increase in the size of the training set.
 
\begin{figure}[htbp]
\centering
\includegraphics[width=0.45\columnwidth]{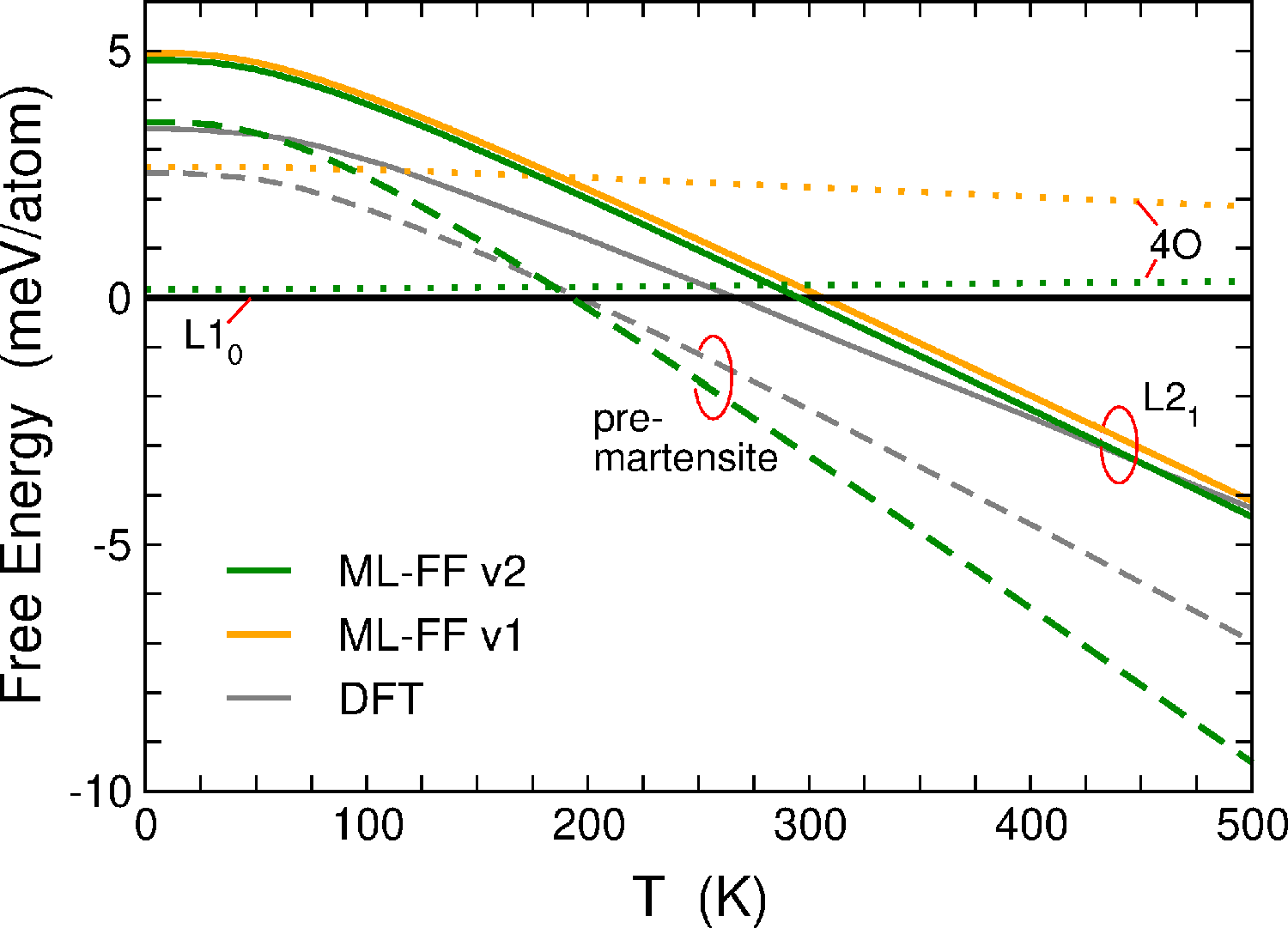}
\caption{
Free energy of \NiMnGa\ from the harmonic approximation calculated with ML-FF v1 (orange lines) and v2 (green lines) for cubic L$2_1$ austenite (solid lines), 3-fold modulated premartensite (dashed lines), 
and 4O martensite (dash-dotted lines) relative to non-modulated L$1_0$ martensite (horizontal black line). The results are compared to previous DFT calculations (grey lines) in the quasi-harmonic approximation, taken from Ref.~\cite{cn:Uijttewaal09}. 
The excellent agreement confirms the transferability of the force field to thermodynamic quantities.}
\label{fig:free_energy}
\end{figure}
Access to the phonon dispersion and the vibrational density of states provides the necessary ingredients to calculate free energies at finite temperatures based on the harmonic approximation,~\cite{cn:Fultz10} which gives a prediction of the martensitic transition 
temperature including quantum statistics at low temperatures and zero-point energies.
In Fig.~\ref{fig:free_energy}, we compare the harmonic free energies of both ML-FF versions for the L$2_1$ austenite, 
4O modulated, and L$1_0$ non-modulated martensite. The latter is taken as an energy reference for each temperature and thus represented as a horizontal line. 
The free energy of the 4O is in general higher than that of the non-modulated martensite, indicating that this modulation does not constitute a stable phase. 
In the case of v2, this is due to the larger zero-point energy. 
The free energies of L$2_1$ and L$1_0$ cross slightly above ambient conditions. This still represents a reasonable agreement with the experimental martensitic transition temperature of $T=240$\,K. 
For ML-FF v2, we also considered the pre-martensite phase, in terms of a 6-fold sinusoidal lattice wave corresponding to the soft TA$_2$ phonon inscribed in the cubic lattice, as a metastable structure. 
This phase was also included in a previous detailed analysis of the free energies in the quasi-harmonic approximation, which additionally includes thermal expansion, based solely on DFT data.~\cite{cn:Uijttewaal09} 
The premartensite is here stabilized with respect to the austenite at $T=0$ by zero-point energy, which arises from particular low-energy modes 
present in this structure. At finite temperatures, this leads to an increasing advantage arising from the entropy contribution, 
which inhibits a crossing between premartensite and L$2_1$, and thus the proper prediction of austenite as a stable phase at higher temperatures. 
In \cite{cn:Uijttewaal09}, this issue was resolved by considering the impact of magnetic interactions on the internal energy and lattice vibrations, 
which leads to excellent agreement with the experimental transition sequence. 
Nevertheless, in our preliminary MD simulations discussed below, we have not yet encountered a stable premartensite phase. 
This disagreement may be due to the omission of quantum statistics in classical MD. Conversely, the coupling between phonon modes at finite temperatures, which can stabilize the soft mode,~\cite{cn:Hickel12} is not captured by the harmonic approximation. 
This calls for a systematic analysis of the vibrational spectrum at finite temperatures in the future. This becomes feasible with the ML-FF developed in this work, given the good overall agreement (within 2~\meVfU) with the DFT free energies from \cite{cn:Uijttewaal09} at all temperatures. 

\subsection{Ultrafast dynamics of laser-heated $\mathrm{Ni_2MnGa}$}
\begin{figure}
  \centering
  \includegraphics[width=0.95\columnwidth]{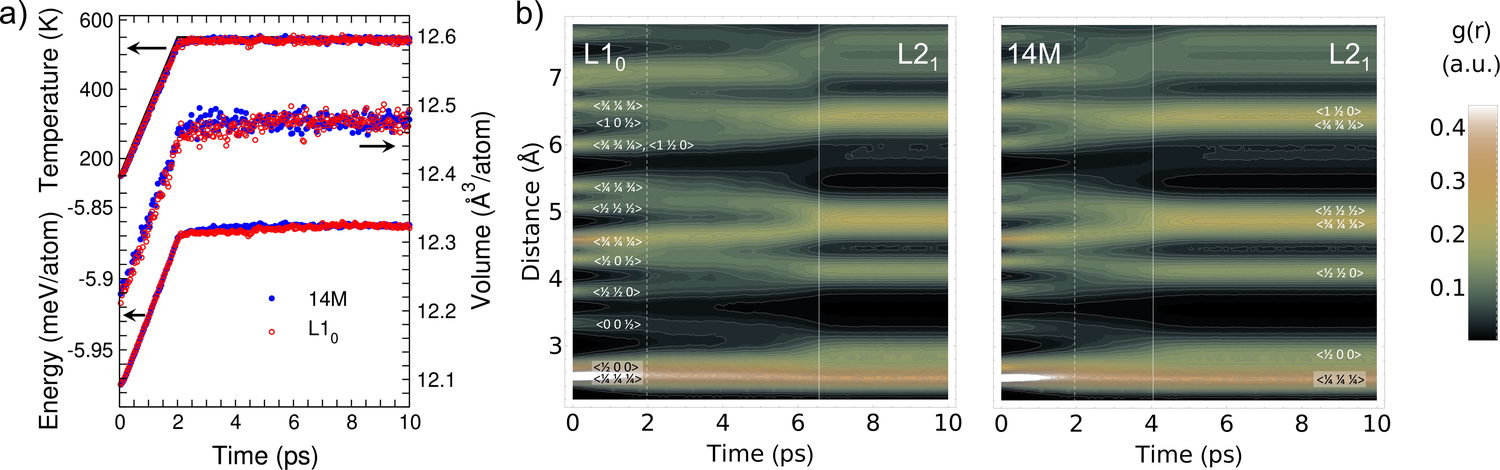}
  \caption{a) Temperature profile of the ultrafast heating process, potential energy per atom, volume of the simulation cell as a function of the simulation time. 
  The black line describes the $NpT$ temperature protocol, 
    raising $T$ from 150 to 550\,K in 2\,ps and holding $T$ afterwards.
    Blue (14M) and red (non-modulated L$1_0$) points denote the actual data for the three quantities measured during the simulation every 20 time steps for the two initial structures. b) 
  Evolution of the pair correlation function $g(r)$ during ultrafast heating of the L1$_0$ (left) and 14M (right). 
    The 14M transforms to cubic L2$_1$ at $t\simeq4$\,ps, while the Bain-path-constrained L1'$_0$ transforms later at $t\simeq6.5$\,ps, indicated by the full vertical white lines. The dashed lines indicate the end of the heating period.
    \label{fig:uheat}}
\end{figure}
To connect the simulations with the ultrafast laser-heating experiments of L\"unser and F\"ahler,~\cite{Ge2025} we performed classical MD simulations in the NpT ensemble to estimate the time scale of the transition on the nm scale, not (yet) taking into account higher-order microstructure.
To mimic the situation where a 270~fs laser pulse first excites the electronic subsystem, which then equilibrates with the lattice within a few picoseconds through electron--phonon coupling, 
we heated a cell consisting of approximately 8000 atoms 
from $T=150$~K to $T=550$~K in 2~ps (corresponding to a heating rate of 200~K/ps) 
and held that temperature for a further 10~ps. 
While the original work was carried out with ML-FF v1, we now discuss simulations with ML-FF v2 for both the non-modulated L$1_0$ martensite and the 14M martensite 
in cells comprising 8\,192 and 10\,752 atoms, respectively. 
The cells are distinguished by the orientation of the primitive cell. 
While for the L$1_0$ the boundaries of the periodically repeated cells are perpendicular to the Cartesian axes 
(i.e., $\{1\,0\,0\}$, $\{0\,1\,0\}$ and $\{0\,0\,1\}$ planes), the upper and lower boundaries of the 14M cells are formed by $\{1\,1\,0\}$ planes instead.
This allows for an easy horizontal shear of those planes, regardless of the cell dimensions.

Fig.~\ref{fig:uheat}a) shows that relevant observables, in particular lattice temperature, volume of the simulation cell, and potential energy, immediately follow the heating and holding protocol, 
with some finite-size-related fluctuations.
Also with ML-FF v2, we obtain characteristic differences between the two cases. The slightly smaller volume per atom of the L$1_0$ structure -- which originates from the absence of twin boundaries -- is compensated by the 14M cell around 4\,ps, while a slight difference in potential energy can be observed up to 6\,ps.
A clearer impression of the differences is given by the contour plot of the pair correlation function $g(r)$ as a function of temperature, which is shown for both structures in
Fig.~\ref{fig:uheat}b). $g(r)$ is proportional to the probability of finding another atom at a given distance $r$ relative to the site of an arbitrary atom. 
The key result is that -- in agreement with the results obtained with ML-FF v1 -- the 14M martensite transforms substantially faster to the cubic L$2_1$ austenite than the L$1_0$ phase.
Specifically, the 14M transforms at $t \approx 4$~ps -- which coincides with the time when the volumes start to coincide -- while the L$1_0$ transforms at $t \approx 6.5$~ps, where the potential energies of both structures converge.
These values differ only slightly from the values published in \cite{Ge2025} for ML-FF v1. 
We therefore conclude that
adaptive nanotwinning with a low $\gamma_{\rm TB}$ profile along the transformation path is advantageous for the martensite$\rightarrow$austenite transition, 
because it facilitates easy shear of $\{1\,1\,0\}$ planes. 
This shear allows the nanotwinned martensite to transform to austenite with smaller atomic displacements compared to the non-modulated L$1_0$. 
In the latter case, the entire cell must transform along the Bain path, explaining the slower transition rate. 

\begin{figure}
  \centering
  \includegraphics[width=0.45\columnwidth]{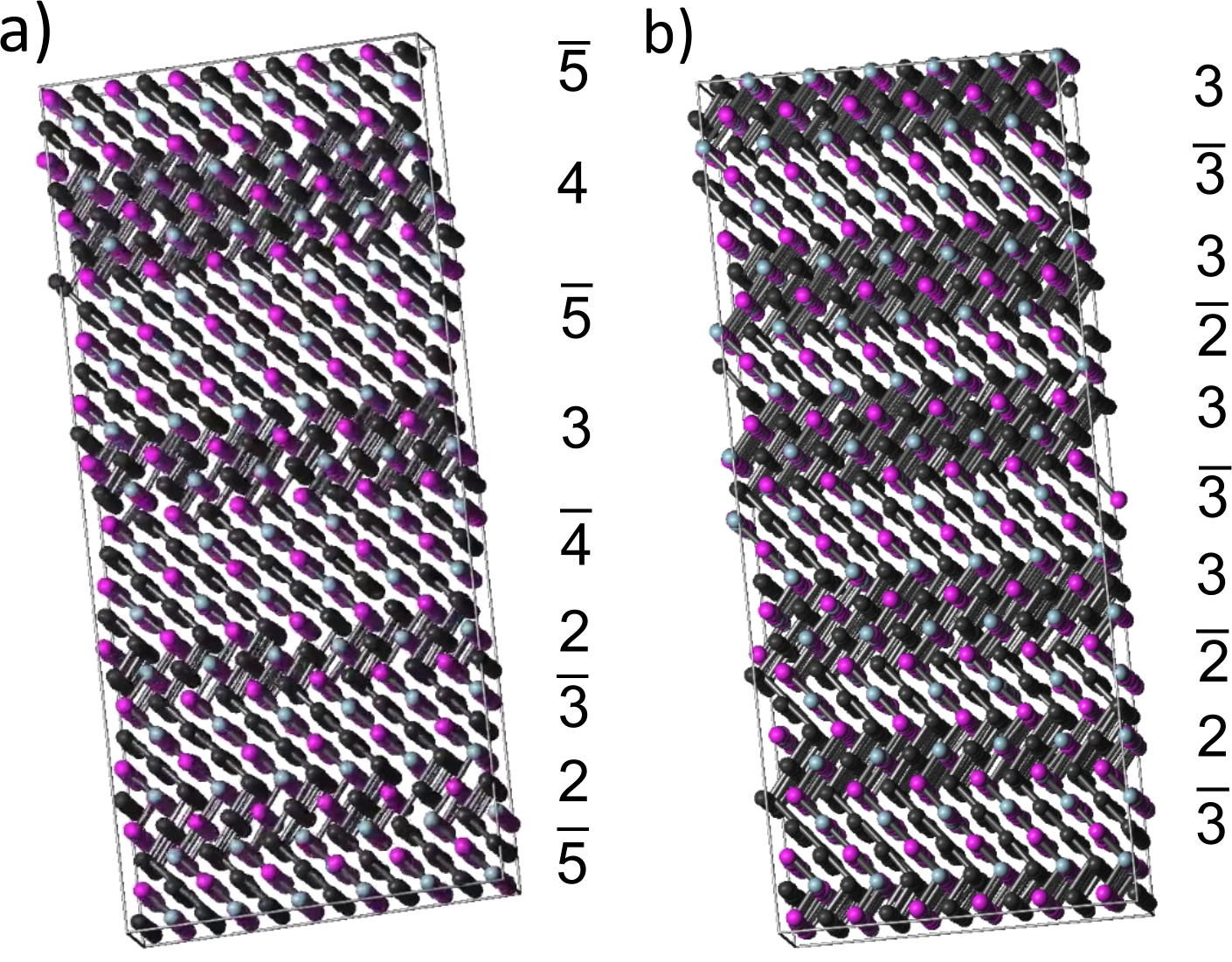}
  \caption{Comparison of martensite structures obtained after cooling down a 4032-atom cell to 150\,K using ML-FF v1 (a) and ML-FF v2 (b). 
  The cells were previously heated to 550\,K and thus in the L$2_1$ austenite state. Black spheres denote Ni atoms, blue Mn and purple Ga.
  Bonds are drawn only along the short $a$- and $b$-axes of the tetragonal martensite, indicating the orientation of the nanotwins. Width and orientation of the individual twins are indicated by the numbers to the right of each structure.
\label{fig:v1v2}}
\end{figure}

For both ML-FF v1 and v2, we cooled previously heated structures of approximately 4000 atoms 
in the L$2_1$ phase from $T=550$~K to $T=150$~K in 200~ps with a linear cooling rate of 2\,K/ps. During cooling, one observes that below room temperature patches of coordinated bonds appear 
in the cell that originally contained the 14M. From these patches, twin domains of an adaptive martensite with an irregular layer sequence evolve. In contrast, the former L$1_0$ cell, which is constrained to transform along the Bain path by the geometry of the cell, remains cubic until the end of the simulation. This demonstrates that adaptive nanotwinning provides a substantial advantage also for the reverse transformation. 
However, for ML-FF v1 and v2, the twin structures evolving in the 14M cell after the phase transition are slightly different, 
as visualized in Fig.~\ref{fig:v1v2}. In the first case, an apparently random arrangement of twins with different widths occurs 
in order to match the periodic boundary conditions, which in our case take over the role of the geometric constraint imposed by the austenite 
at the phase interface in Khachaturyan's concept of adaptive martensite~\cite{cn:Khachaturyan91,cn:Kauffmann11}. 
In the case of ML-FF v2, the simulation ends up in a more regular state, consisting mainly of 2- and 3-layer twins. 
We ascribe this difference to the negative twin boundary energy $\gamma_{\rm TB}$ captured by ML-FF v2 (but not by v1), which energetically favors structures with a high density of twin boundaries. 
This comparison between v1 and v2 suggests that the degree of nanotwin ordering in the final martensite is directly controlled by $\gamma_{\rm TB}$.

\subsection{Slow heating and cooling of nanotwinned and non-modulated martensite} 
\begin{figure}[htbp]
\centering
\includegraphics[width=0.95\columnwidth]{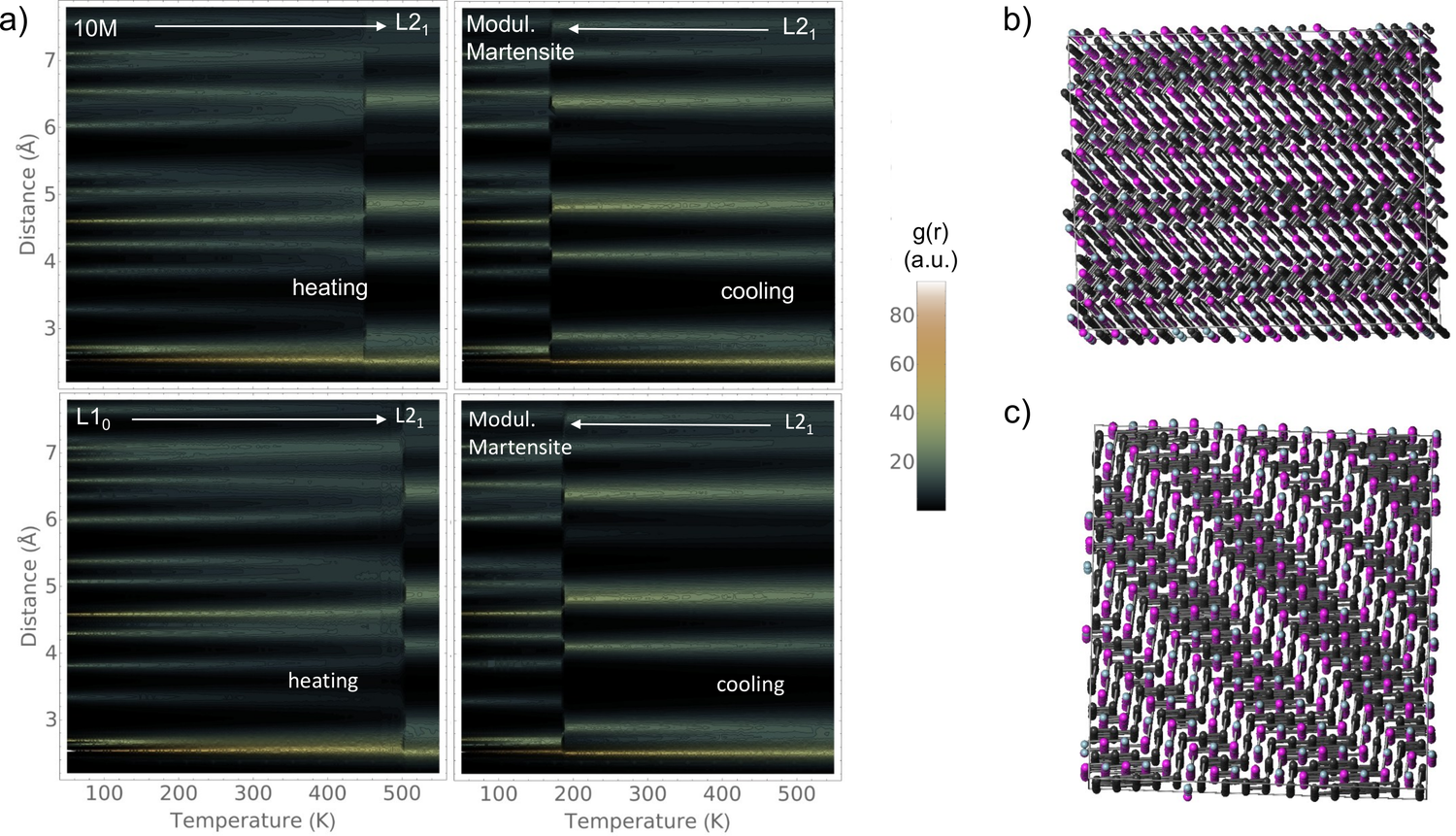}
\caption{a) Time evolution of the pair correlation function $g(r)$ of a 10M (upper panels) and L$1_0$ martensite (lower panels) during heating (left column) and cooling (right column) 
with a cooling rate of $0.25$\,K/ps.  b) and c) Final structures obtained after cooling for initial 10M and L$1_0$, respectively.}
\label{fig:gr_slow}
\end{figure}
Since the nucleation of another phase in a first-order transition is a stochastic process, 
substantially slow heating and cooling rates are required 
for a reasonably accurate screening of the transition temperatures within classical MD. We therefore repeated the simulations for linear heating and subsequent cooling with a rate of $0.25\,$K/ps, 
which corresponds to a total simulation time of 2\,ns. We carried this procedure out for the 
non-modulated L$1_0$, 14M, 6M, and 10M.
The corresponding pair distribution functions for heating and cooling of cells containing the 10M (7\,680 atoms) and L$1_0$ (8192 atoms) are shown in Fig.~\ref{fig:gr_slow}a). Both structures undergo forward and backward transitions, however at different temperatures. 
Upon heating, the 10M transforms to the cubic L$2_1$ phase around $T=450\,$K, 
and the non-modulated L$1_0$ at a substantially higher temperature, around 500\,K, which possibly indicates a higher energy barrier to be overcome by a transition along the Bain path. The reverse transformation occurs at a substantially lower temperature, around $T=170\,$K for the former 10M and 185\,K for the L$1_0$, which are much closer. In fact, the latter structure does not return to the non-modulated L$1_0$ but evolves to a nanotwinned arrangement with boundaries along $[1\,0\,1]$ (see Fig.~\ref{fig:gr_slow}c).
The final configuration of the other cell reminds of the original 10M, although the nanotwin boundaries do not appear to be perfectly aligned throughout the complete cell. 
The slightly earlier transition in the L$1_0$ cell may result from the cubic lattice symmetry of the L$2_1$ austenite,
which allows for 6 independent possibilities to introduce twin interfaces in the cell, potentially increasing the probability for a suitable nucleation event. The transition temperatures of the other modulated structures (not shown) vary slightly but are in general closer to the values obtained for the 10M structure. 

\section{Conclusions and outlook}
\label{sec:conclusion}
In this work, we provide a proof of concept that machine-learning force fields trained on first-principles data constitute a suitable tool 
to model martensitic transformations in the stoichiometrically ordered magnetic shape memory alloy Ni$_2$MnGa, which is characterized by intricately linked structural, electronic, and magnetic instabilities. 
This opens a new avenue towards a parameter-free modelling of the functional properties of this system at finite temperatures. 
We verified that, after training on 3074 reference configurations, the most relevant features of the total energy known from previous DFT work are reproduced within an error of 2~\meVfU. 
This also holds for free energy differences up to 500\,K, while the force constants can reproduce the entire phonon dispersion of austenite, including the imaginary TA$_2$ mode along $\Gamma$--K. 
Although only ferromagnetic configurations entered the training set, 
the ML-FF obtained in this way is capable of simulating 
austenite, non-modulated, and nanotwinned martensites 
including a forward and backward martensitic transformation, providing a first realistic atomistic model that can be used efficiently at finite temperatures. 
However, we encounter a substantial hysteresis, which we partially ascribe to the absence of defects, disorder, off-stoichiometric compositions, and magnetic fluctuations in the training set.

The controlled comparison between ML-FF v1 (baseline, without nanotwinned intermediate states) and ML-FF v2 (with intermediate martensites) demonstrates that the training data representing the nanotwinned intermediate martensites are the decisive ingredient for reaching negative twin boundary energy $\gamma{\rm TB}<0$ over a wide range of intermediate martensitic distortions. Additional training of the austenite and premartensite close to $T=0$ restores the imaginary TA$2$ phonon branch along $\Gamma$--K. But even the baseline v1, with $\gamma_{\rm TB}\approx 0$ for small tetragonal distortions, already supports the easy shear of $\{1\,1\,0\}$ planes. In the simulated ultrafast heating experiment this leads to a substantially faster transformation of the 14M modulated martensite compared to the non-modulated L$1_0$ phase, which is forced to transform along the Bain path, imposing longer distances for the ions to move. 
The negative $\gamma_{\rm TB}<0$ captured by v2 further promotes the formation of an approximately ordered nanotwin configuration during cooling.

Two limitations constrain the quantitative accuracy of the present force field. First, the training set contains only stoichiometric, fully ferromagnetic configurations. 
In our view, the substantial hysteresis observed in the slow heating and cooling simulations could arise from an overestimated nucleation barrier connected to the absence of chemical and magnetic disorder.
 Future work should thus extend the training set to include off-stoichiometric compositions (e.g., Ni$_{2+x}$Mn$_{2-x}$Ga), magnetic disorder, and interfacial structures,
 and move to simulations with free surfaces or realistic defects.
Second, inaccuracies also arise from the technical settings of the underlying DFT training set, which is often limited by computational efficiency. For truly quantitative predictions, an accuracy of less than a \meVfU{} will be required.
 As recently demonstrated in~\cite{cn:Zeleny21}, the DFT+U approach might offer a practical route to improve the agreement between the DFT reference data and experiment.
Since numerical complexity increases with any improvement, systematic benchmarking protocol against both DFT and experimental data (elastic constants, transformation temperatures, twin boundary mobility) will be necessary to establish quality criteria before applying the force field to quantitative predictions.

In summary, we have shown that a DFT-trained ML-FF can reproduce the key $T=0$ features of \NiMnGa{} identified in previous first-principles work---the soft TA$_2$ phonon along $\Gamma$--K, the stability of nano-twinned intermediate martensites with $\gamma_{\rm TB}\leq 0$ leading to the close energetic competition between NM, 4O, and 10M martensites. In this way, we extended atomistic simulations to 8000--10\,000 atoms over nanosecond time scales -- a regime that is computationally inaccessible to pure first-principles methods. The controlled comparison between ML-FF v1 (without training data involving intermediate nano-twinned martensites) and ML-FF v2 (containing those) indicates that negative twin boundary energies are a decisive ingredient for capturing the transformation into the proper modulated martensites.
This provides a direct atomistic confirmation of the kinetic prediction of the adaptive-martensite framework and establishes a foundation for future large-scale modelling of functional properties in magnetic shape memory alloys at finite temperatures.

\section*{Acknowledgments}
M.E.G.\ would like to thank Sebastian F\"ahler (HZDR, Dresden) and Klara L\"unser (University of Duisburg-Essen) for intensive discussions on the dynamics of martensitic transformations and Petr Jaros, Petr Sedl\'ak (Charles University, Prague), Martin Zelen\'y and Petr \v{S}est\'ak (Brno University of Technology) for important exchange regarding the generation and validation of ML-FF for shape memory alloys. The authors thank the German Science Foundation (DFG) for financial support through
 TRR270, project-ID 405553726, subprojects B06, B01, A10 (M.E.G., Y.K. K.P.S.) and CRC1242 project no. 278162697, subproject C02 (M.E.G., M.J.B). O.N.M. acknowledges support by the DFG through project MI 3273/1. Calculations were partially carried out on the high performance computer systems of the UDE “MagnitUDE” and “AmplitUDE” (DFG inst 20876/209-1 and 20876/243-1 FUGG) and the Lichtenberg~II High Performance Computer at the NHR Center NHR4CES at TU Darmstadt (project number p0020039).

\section*{Data Availability Statement}

Inputs, results and processed data of the MD simulations, including the force fields are available at https://doi.org/url\_to\_be\_inserted\_upon\_publication. 

\bibliographystyle{apsrev4-2}
\bibliography{refs}

\end{document}